\documentclass[dvips]{article}
\usepackage{supertabular,lscape,epsfig}
\usepackage{amssymb}
\usepackage{amsmath}

\DeclareSymbolFont{ppa}{OT1}{ppl}{m}{it}
\DeclareMathSymbol{\vv}{\mathalpha}{ppa}{'166}

\begin{document}

\newcommand{\dd}{\,{\rm d}}
\newcommand{\ie}{{\it i.e.},\,}
\newcommand{\etal}{{\it et al.\ }}
\newcommand{\eg}{{\it e.g.},\,}
\newcommand{\cf}{{\it cf.\ }}
\newcommand{\vs}{{\it vs.\ }}
\newcommand{\zdot}{\makebox[0pt][l]{.}}
\newcommand{\up}[1]{\ifmmode^{\rm #1}\else$^{\rm #1}$\fi}
\newcommand{\dn}[1]{\ifmmode_{\rm #1}\else$_{\rm #1}$\fi}
\newcommand{\upd}{\up{d}}
\newcommand{\uph}{\up{h}}
\newcommand{\upm}{\up{m}}
\newcommand{\ups}{\up{s}}
\newcommand{\arcd}{\ifmmode^{\circ}\else$^{\circ}$\fi}
\newcommand{\arcm}{\ifmmode{'}\else$'$\fi}
\newcommand{\arcs}{\ifmmode{''}\else$''$\fi}
\newcommand{\MS}{{\rm M}\ifmmode_{\odot}\else$_{\odot}$\fi}
\newcommand{\RS}{{\rm R}\ifmmode_{\odot}\else$_{\odot}$\fi}
\newcommand{\LS}{{\rm L}\ifmmode_{\odot}\else$_{\odot}$\fi}

\newcommand{\Abstract}[2]{{\footnotesize\begin{center}ABSTRACT\end{center}
\vspace{1mm}\par#1\par
\noindent
{~}{\it #2}}}

\newcommand{\TabCap}[2]{\begin{center}\parbox[t]{#1}{\begin{center}
  \small {\spaceskip 2pt plus 1pt minus 1pt T a b l e}
  \refstepcounter{table}\thetable \\[2mm]
  \footnotesize #2 \end{center}}\end{center}}

\newcommand{\TableSep}[2]{\begin{table}[p]\vspace{#1}
\TabCap{#2}\end{table}}

\newcommand{\FigCap}[1]{\footnotesize\par\noindent Fig.\  %
  \refstepcounter{figure}\thefigure. #1\par}

\newcommand{\TableFont}{\footnotesize}
\newcommand{\TableFontIt}{\ttit}
\newcommand{\SetTableFont}[1]{\renewcommand{\TableFont}{#1}}

\newcommand{\MakeTable}[4]{\begin{table}[htb]\TabCap{#2}{#3}
  \begin{center} \TableFont \begin{tabular}{#1} #4 
  \end{tabular}\end{center}\end{table}}

\newcommand{\MakeTableSep}[4]{\begin{table}[p]\TabCap{#2}{#3}
  \begin{center} \TableFont \begin{tabular}{#1} #4 
  \end{tabular}\end{center}\end{table}}

\newenvironment{references}%
{
\footnotesize \frenchspacing
\renewcommand{\thesection}{}
\renewcommand{\in}{{\rm in }}
\renewcommand{\AA}{Astron.\ Astrophys.}
\newcommand{\AAS}{Astron.~Astrophys.~Suppl.~Ser.}
\newcommand{\ApJ}{Astrophys.\ J.}
\newcommand{\ApJS}{Astrophys.\ J.~Suppl.~Ser.}
\newcommand{\ApJL}{Astrophys.\ J.~Letters}
\newcommand{\AJ}{Astron.\ J.}
\newcommand{\IBVS}{IBVS}
\newcommand{\PASP}{P.A.S.P.}
\newcommand{\Acta}{Acta Astron.}
\newcommand{\MNRAS}{MNRAS}
\renewcommand{\and}{{\rm and }}
\section{{\rm REFERENCES}}
\sloppy \hyphenpenalty10000
\begin{list}{}{\leftmargin1cm\listparindent-1cm
\itemindent\listparindent\parsep0pt\itemsep0pt}}%
{\end{list}\vspace{2mm}}

\def\TYLDA{~}
\newlength{\DW}
\settowidth{\DW}{0}
\newcommand{\dw}{\hspace{\DW}}

\newcommand{\refitem}[5]{\item[]{#1} #2%
\def\REFARG{#3}\ifx\REFARG\TYLDA\else, {\it#3}\fi
\def\REFARG{#4}\ifx\REFARG\TYLDA\else, {\bf#4}\fi
\def\REFARG{#5}\ifx\REFARG\TYLDA\else, {#5}\fi.}

\newcommand{\Section}[1]{\section{\hskip-6mm.\hskip3mm#1}}
\newcommand{\Subsection}[1]{\subsection{#1}}
\newcommand{\Acknow}[1]{\par\vspace{5mm}{\bf Acknowledgements.} #1}
\pagestyle{myheadings}

\newfont{\bb}{ptmbi8t at 12pt}
\newcommand{\xrule}{\rule{0pt}{2.5ex}}
\newcommand{\xxrule}{\rule[-1.8ex]{0pt}{4.5ex}}
\def\thefootnote{\fnsymbol{footnote}}
\begin{center}
{\Large\bf Binary Lenses in OGLE-II 1997--1999 Database.~
      A Preliminary Study}
\vskip.6cm
{\bf M.~~ J~a~r~o~s~z~y~\'n~s~k~i}
\vskip2mm
{Warsaw University Observatory, Al.~Ujazdowskie~4,~00-478~Warszawa, Poland\\
e-mail:mj@astrouw.edu.pl}
\end{center}

\vspace*{7pt} 
\Abstract{We present 18 binary lens candidates from OGLE-II database for 
seasons 1997--1999. The candidates have been selected by visual light curves 
inspection from the subsample of strong transient events; the same procedure 
gives 215 single lens candidates. Among the double lenses there are 12 cases 
interpreted as caustic crossing events. We compare the mass ratio and 
separation distributions obtained for binary lenses with the predictions based 
on stellar double systems observations. We take into account the selection 
bias, which causes over-representation of binary lenses of similar mass and 
separation close to the Einstein radius. There is no strong discrepancy 
between the expected and observed distributions of the mass ratio or the 
binary separations. We find one or two cases of binary lens candidates, 
SC20\_1793 and SC20\_3525, with extreme mass ratios, which may suggest 
presence of planets or brown dwarf companions. Unfortunately, neither case is 
very strong, as alternative solutions provide fits to the data which are only 
unsubstantially worse. Binary lenses provide a modest contribution to overall 
optical depth to microlensing.}{galaxies: individual (Milky Way) -- 
gravitational lensing
}{}

\Section{Introduction}
As estimated by Mao and Paczy\'nski (1991) several percent of all microlensing 
events in our Galaxy should be caused by binaries acting as lenses. The first 
such case, the event OGLE-7 (Udalski \etal 1994a) was  analyzed also by 
Dominik (1999), who showed the existence of several different binary lens 
models of the event. 

The amount of information one may obtain from a binary lens light curve 
depends strongly on the frequency of observations and the kind of the event 
itself. The events with well sampled caustic crossing may be used to study the 
effect of limb darkening in the atmospheres of lensed stars (\eg Albrow \etal 
1999b) or to set limits on the source velocity relative to the lens (\eg 
Albrow \etal 1999a). In some cases with caustic crossings and a pronounced 
effect of  Earth motion parallax, the mass of the lens can be found (An \etal 
2001). Such finding is not possible for single gravitational lenses and rather 
uncommon for doubles. The typical caustic crossing takes several hours, and 
the observations should be taken every ten or twenty minutes to be useful for 
detailed modeling. The standard frequency of observations used by the 
gravitational lens survey teams (including OGLE-II -- Udalski \etal 2000) is 
one or two per night -- insufficient to treat the caustic crossings as 
separate issue. Some properties of binary lenses can be found even without 
well sampled caustic crossings. 

The only systematic study to date of the binary lens population was made for 
the MACHO data (Alcock \etal 2000). It gives the light curves and binary lens 
models for 21 events observed in 1993--1998. Many of the events were known to 
be caused by binary lenses early enough to make the follow up observations 
with the dense sampling of the caustic crossing. In four cases the relative 
proper motion of the lens was measured with the accuracy sufficient to study 
the likelihood of the lens masses. The distribution of the binary mass ratio 
is also given. 

Wo\'zniak \etal (2001) present a list of 520 gravitational lens candidates among 
OGLE-II stars based on the new data processing using DIA photometry (Alard and 
Lupton 1998, Alard 2000). The database includes observations of seasons 
1997--1999. Most of the candidates have been found algorithmically. The list 
is supplemented by candidates selected by visual inspection of the light 
curves and also by some events which were known as candidates in earlier study 
(Udalski \etal 2000). The sample of the gravitational lens candidates belongs 
to a larger sample of transient sources. The transient sample (Wo\'zniak 2000, 
Wo\'zniak \etal 2001) has been selected in fully algorithmic manner from all 
OGLE-II light curves and serves as the basis of the present study. 

Our goal is to find the candidates for binary lens events, find their models 
and make some survey of the binary lens population properties. In the next 
Section we describe our selection method for finding both single and binary 
lens candidates in the transient sample of Wo\'zniak \etal (2001). In Section~3 
we describe the procedure of fitting binary lens models to the data. The 
results are described in Section~4, while the graphical material is shown in 
Appendix. Discussion follows in Section~5. 

\Section{Choice of Candidates}
Wo\'zniak (2000) and Wo\'zniak \etal (2001) describe the construction of the 
transient sample based on OGLE-II data for seasons 1997--1999 processed using 
the DIA photometry. The sample contains 4424 sources. The transients are 
selected algorithmically, as sources which have (almost) constant flux of 
radiation and undergo at most two brightening episodes during observation 
span. We follow Wo\'zniak in defining a threshold value for physical flux 
variation. We assume that the variability of the source is limited to a 
sequence containing not more than 50\% of all observations. The remaining data 
points serve to define a constant flux base. The choice of the ``variable'' 
part is made in  a way minimizing the scatter of the remaining observations 
(the ``base''). We calculate $\sigma_{\rm base}$ -- a {\it rms} deviation of 
flux measurements from its mean value $F_{\rm base}$. Following Wo\'zniak (2000) 
we adjust the calculated parameter taking into account possibly non-Gaussian 
distribution of flux values around the mean. Finally, we obtain the threshold 
parameter ${\delta=\max\{\sigma_{\rm base},\sigma_{\rm ph}\}}$, where 
$\sigma_{\rm ph}$ is the mean photometric error estimated for the points 
belonging to the constant flux database. Wo\'zniak \etal (2001) define a 
brightening episode as 3 consecutive points at least $5\delta$ above the base 
flux or 4 points at least $4\delta$ above. 

The light curves of binary lens events usually have two maxima and it is 
necessary to have at least five measurements to distinguish them. For this 
reason and having in mind some safety margin, we require the transients to 
have at least 7 points which are $5\delta$ above the base flux. Some 
experience with Wo\'zniak \etal (2001) database suggests an extra criterion for 
selecting the interesting events, which helps removing DIA photometry 
artifacts from the investigated light curves. In further investigation we 
include only the light curves with at least 7 points which are $5\delta$ above 
the base flux with at least one point among them at $15\delta$ level. 

Using our criteria we automatically search the transient database getting 390 
light curves. We visually inspect all the selected cases. For each light curve 
we try a single lens fit taking into account blending. In the majority of cases 
comparison of the best single lens fit with the data shows a good agreement, 
in the sense of visual impression and the $\chi^2$ value. There are cases of 
weak transients, which can be formally fitted with single lens model based on 
$\chi^2$ value, which are nevertheless rejected after visual inspection,
usually the changes in flux look as random with a period of increased mean 
flux value. There are also cases which under visual inspection show a very 
good agreement with a single lens model light curve, but have high $\chi^2$ 
value. Usually these are the strongest lensing events. We treat them as single 
lens events not investigating further the reasons for low formal quality of 
the fits. The second most common category among selected transients are 
artifacts of DIA photometry. They show irregular behavior dominated by noise. 
In many cases series of sources close to each other show very similar light 
curves of this kind. In a few cases the light curves suggest periodic 
variability. We exclude the possibility that there are binary lenses belonging 
to this category. The binary lens candidates are chosen mostly on the basis of 
visual inspection. Some show the characteristic ``U-shape'' of the caustic 
crossing. Some have multiple flux maxima. The other show asymmetry. All 
visually selected candidates have high values of $\chi^2$ for a single lens 
model, usually exceeding 3 per degree of freedom. There are also several cases 
of good quality light curves, which do not resemble a single or binary lens 
model variability. We put them into ``other objects'' category. 

Among 390 transient selected we find 215 single lenses and 141 unclassifiable 
cases of poor photometry. We treat 27 of the remaining 34 light curves as 
possible binary lens candidates, and 7 -- as ``other objects''. 

\Section{Fitting Binary Lens Models}
The models of the two point mass lens were investigated by many authors 
(Schneider and Weiss 1986, Mao and DiStefano 1995, DiStefano and Mao 1996, 
Dominik 1998, to mention few). The methods applicable for extended sources 
have recently been described by Mao and Loeb (2001). While we use mostly the 
point source approximation, we extensively employ their efficient numerical 
schemes for calculating the binary lens caustic structure and source 
magnification. 

We fit binary lens models using the $\chi^2$ minimization  method for the 
light curves. It is convenient to model the flux at the time $t_i$ as: 
$$F(t_i)=F_b+F_s\times A(t_i)\equiv F_{\rm base}+F_s\times\left(A(t_i)-1\right)\eqno(1)$$
where $F_b$ is the blended flux (from the source close neighbors and possibly 
the lens), $F_s$ -- the flux of the source which is lensed, and $A(t_i)$ is the 
lens magnification. The combination ${F_{\rm base}\equiv F_b+F_s}$ is the 
total flux long before or long after the event. The DIA photometry only gives 
the flux differences, so one can use ${\tilde F_i\equiv F_i-F_{\rm base}}$ as 
``measured'' quantities. Using this notation one has for the $\chi^2$ (Wo\'zniak 
2000): 
$$\chi^2=\sum_{i=1}^N\frac{\left((A_i-1)F_s-\tilde F_i\right)^2}{\sigma_i^2}\eqno(2)$$
where $\sigma_i$ are the errors of the flux measurement taken from the DIA 
photometry. For self-consistency we check the value of $\chi^2$ calculated for 
the points defining the constant flux base.

Binary lens models are possibly and typically non unique (Dominik 1999). The 
presence of caustics and cusps in the lens theory (Schneider, Ehlers and Falco 
1992, Blandford and Narayan 1992) makes the $\chi^2$ dependence on the model 
parameters complicated and discontinuous for point sources. This suggests that 
the search for the $\chi^2$ global minimum may be difficult because of the 
likely existence of many local ones. For this reason we combine the scan of 
the parameter space with the standard minimization techniques in our search. 

The light curves we investigate do not have well sampled caustic crossings, so 
we use the point source model in the majority of calculations. For the same 
reason we are also not able to use the strategy of Albrow \etal (1999c) for 
finding binary lens models. We also neglect the parallax and/or binary 
rotation effects, which further simplifies the models. 

The binary system consists of two masses $m_1$ and $m_2$, where by convention 
${m_1\le m_2}$. The Einstein radius of the binary lens is defined as: 
$$r_{\rm E}=\sqrt{\frac{4G(m_1+m_2)}{c^2}\frac{d_{\rm OL}d_{\rm LS}}
{d_{\rm OS}}}\eqno(3)$$
where $G$ is the constant of gravity, $c$ is the speed of light, $d_{\rm OL}$ 
is the observer--lens distance, $d_{\rm OS}$ is the observer--source 
distance, and $d_{\rm LS}$ is the distance between the lens and the source. 
The Einstein radius serves as a length unit and the Einstein time: ${t_{\rm E} 
=r_{\rm E}/\vv_\perp}$, where $\vv_\perp$ is the lens velocity relative to the 
line joining the observer with the source, serves as a time unit. The passage 
of the source in the lens background is defined by seven parameters: ${q\equiv 
m_1/m_2}$ (${0<q\le1}$) -- the binary mass ratio, $d$ -- binary separation 
expressed in $r_{\rm E}$ units, $\beta$ -- the angle between the source 
trajectory as projected onto the sky and the projection of the binary axis, 
$b$ -- the impact parameter, $t_0$ -- the time of closest approach of the 
source to the binary center of mass, $t_E$ -- the Einstein time, and $F_s$ the 
(not magnified) source flux. For all other parameters fixed, $F_s$ can be 
obtained from the (linear) equation ${\partial\chi^2/\partial F_s=0}$ with the 
obvious limitation ${0\le F_s\le F_{\rm base}}$. Thus we are left with the 
six dimensional parameter space to look for minima. We begin with a scan of 
the parameter space using a logarithmic grid of points in $(q,d)$ plane 
(${10^{-3}\le q\le1}$, ${0.1\le d\le10}$) and allowing for 
continuous variation of the other four parameters (inclusion of $q$ and $d$, 
which change the caustic structure of the binary, in the automatic routine 
finding  $\chi^2$ minimum may cause problems with its convergence -- see An 
\etal (2001) Jaroszy\'nski and Mao (2001)). The other dimensions are searched 
using AMOEBA routine (Press \etal 1992). We use many starting points in 
$(\beta,b,t_0,t_E)$ subspace. The expected value for the Einstein time 
$t_E$ is based on the event duration. The expected time of the closest 
approach to the binary center of mass $t_0$ should be close to the time of the 
maximal observed flux. The direction of the source motion relative to the 
binary axis $\beta$ is chosen at random. For each choice of $\beta$ there is a 
defined range of impact parameter $b$ giving rise to a ``binary character'' of 
the light curve. Practically it means that the source should either cross a 
caustic  or pass not too far from it.

Using the point source approximation we find the best binary lens models for 
every $(q,d)$ on the grid. With the fitted value of the Einstein time $t_{\rm 
E}$ and the typical velocity of a source relative to the observer--lens line 
$\vv$, we are able to estimate the linear size of the Einstein radius as 
projected into the source plane ${\tilde r_{\rm E}\approx \vv t_{\rm E}}$. The 
fitted deblended source flux allows for a crude estimate of the star radius 
$R_s$, and corresponding dimensionless source radius ${r_s\equiv R_s/{\tilde 
r_{\rm E}}}$. For self consistency we repeat the fitting procedure at every 
grid point using the extended source model of the estimated size and starting 
from the solution obtained in the point source approximation. Finally we 
search the regions of $(q,d)$ plane with the lowest values of $\chi^2$ on a 
finer grid. 

\Section{Results}
We have applied the fitting procedure to the 27 selected events. We also 
visually compare the top quality models with the data, checking whether the 
fitted light curves are ``natural'', not having superficial peaks in places 
not covered by observations. In eighteen cases the binary lens fits are 
formally good and natural. They are also substantially better than single lens 
fits applied to the same data (the binary fits give the $\chi^2$ values lower 
at least by 50, and typically by several hundreds or more as compared with 
single lens fits.) There are three candidates with formally good fits which 
are rejected after the visual inspection of the models. The other six are 
rejected because of the high $\chi^2$ value. The plots showing light curves 
and our best fits for the rejected candidates are included in Appendix.

Nine of the events are on the Udalski \etal (2000) list of microlensing 
events, and five of them were detected in real time and included in the alerts 
of the Early Warning System (Udalski \etal 1994b). One of the events 
(SC21\_6195) has been included in Alcock \etal (2000) study of binary lenses 
as candidate event 97-BLG-d2. Not all events marked as possible binary lens 
candidates by Udalski \etal (2000) are on our list. We give the comparison of 
the two lists in the Section~5.

\subsection{The Binary Lens Models}
Of the eighteen binary lens candidates, twelve show caustic crossing, five at  
least one close cusp approach with sharp light maximum, and one has two 
shallow maxima explained by passages at greater distances from cusps. 

\MakeTable{lrccccc}{12.5cm}{Binary microlensing events}
{\hline
\noalign{\vskip3pt}
\multicolumn{4}{c}{Identification}& RA & DEC & $I_0$ \\
Field & DIA & OGLE & EWS & (2000) & (2000) & [mag] \\ 
\noalign{\vskip3pt}
\hline
\noalign{\vskip3pt}
BUL\_SC3 & 7390 & 792295 & 1998-BUL-28 & 17\uph53\upm54\zdot\ups14 & $-29\arcd36\arcm40\zdot\arcs8$ & 19.461 \\
BUL\_SC4 & 6350 &- & - &                 17\uph54\upm47\zdot\ups47 & $-29\arcd34\arcm19\zdot\arcs4$ & 19.229 \\
BUL\_SC5 & 6550 & - & - &                17\uph50\upm51\zdot\ups52 & $-29\arcd34\arcm11\zdot\arcs6$ & 16.177 \\
BUL\_SC12 & 997 & - & - &                18\uph15\upm59\zdot\ups02 & $-24\arcd10\arcm04\zdot\arcs8$ & 19.605 \\
BUL\_SC15 & 1631 & 373196 & - &          17\uph48\upm07\zdot\ups05 & $-23\arcd12\arcm03\zdot\arcs2$ & 19.527 \\
BUL\_SC16 & 1047 & 32304 & - &           18\uph09\upm38\zdot\ups47 & $-26\arcd32\arcm26\zdot\arcs8$ & 18.422 \\
BUL\_SC18 & 4924 & - & - &               18\uph06\upm54\zdot\ups66 & $-26\arcd54\arcm02\zdot\arcs8$ & 18.541 \\
BUL\_SC19 & 668  & 26606 & 1999-BUL-23 & 18\uph07\upm45\zdot\ups14 & $-27\arcd33\arcm15\zdot\arcs2$ & 17.868 \\
BUL\_SC20 & 1793 & - & - &               17\uph59\upm02\zdot\ups69 & $-29\arcd03\arcm03\zdot\arcs0$ & 15.377 \\
BUL\_SC20 & 3525 & 708586 & 1999-BUL-25 &17\uph59\upm41\zdot\ups14 & $-28\arcd47\arcm18\zdot\arcs4$ & 18.329 \\
BUL\_SC20 & 4694 & - & - &               17\uph59\upm14\zdot\ups40 & $-28\arcd36\arcm55\zdot\arcs6$ & 19.083 \\
BUL\_SC21 & 6195 & 833776 & (97-BLG-d2)& 18\uph00\upm39\zdot\ups58 & $-28\arcd34\arcm43\zdot\arcs3$ & 17.721 \\
BUL\_SC30 & 4491 & - & - &               18\uph01\upm44\zdot\ups29 & $-28\arcd41\arcm30\zdot\arcs0$ & 18.576 \\
BUL\_SC31 & 1795 & 293442 & 1999-BUL-17 &18\uph02\upm22\zdot\ups50 & $-28\arcd40\arcm43\zdot\arcs9$ & 18.638 \\
BUL\_SC31 & 3204 & - & - &               18\uph02\upm17\zdot\ups56 & $-28\arcd24\arcm10\zdot\arcs1$ & 19.643 \\
BUL\_SC32 & 4683 & - & - &               18\uph02\upm57\zdot\ups13 & $-28\arcd12\arcm53\zdot\arcs3$ & 17.241 \\
BUL\_SC35 & 2526 & 305604 & - &          18\uph04\upm22\zdot\ups42 & $-27\arcd57\arcm52\zdot\arcs2$ & 18.905 \\
BUL\_SC36 & 4030 & 336761 & 1999-BUL-11 &18\uph05\upm29\zdot\ups59 & $-27\arcd59\arcm14\zdot\arcs3$ & 18.537 \\ 
\noalign{\vskip3pt}
\hline
\noalign{\vskip3pt}
\multicolumn{7}{p{11.5cm}}{Note: The event identification consists of the OGLE 
field number, the star number in DIA database, and optionally of OGLE star 
number and EWS event name. The other columns give the star position ($\alpha$, 
$\delta$), and the base magnitude in the $I$-band.}} 

In Table~1 we give the names, positions and observed magnitudes in $I$ band of 
the objects long before or long after the event. 

In Table~2 we give the parameters for the best models of each of the events. 
In some cases we include also another solution if it belongs to a distinct 
$\chi^2$ minimum and lies inside the confidence region. Our best fits are also 
shown on the plots in Appendix. For each binary there are three separate 
plots. The first shows the source trajectory as projected onto the lens plane 
with caustic structure and binary components included. The model light curves 
and observed fluxes are shown in the second plot. The third diagram shows the 
$\chi^2$ confidence regions in the $\lg q$--$\lg d$ plane. 
\MakeTable{lrcccrrrrcc}{12.5cm}{The binary lens models parameters}
{\hline
\noalign{\vskip3pt}
Field & DIA & $\chi^2/$d.o.f. & $q$ & $d$ & $\beta~~~$ & $b~~~$ & $t_0~~~$ &  $t_E~~~$ & $f$  & $r_s$ \\
\noalign{\vskip3pt}
\hline
\noalign{\vskip3pt}
SC3 & 7390  & 309.6/288 & 0.452 & 0.376 &  132.07 & $ 0.10$ &  1067.4 &  113.5 &  0.99 & 0.0006 \\
SC4 & 6350  & 259.4/272 & 0.306 & 0.902 &  115.82 & $-0.30$ &  1376.3 &   71.3 &  0.36 & 0.0006 \\
    &       & 261.6/272 & 0.631 & 1.259 &  126.15 & $ 0.07$ &  1376.2 &   66.2 &  0.27 & 0.0005 \\
SC5 & 6550  & 278.3/272 & 0.966 & 0.881 &   76.05 & $ 0.01$ &   615.2 &   24.6 &  0.01 & 0.0011 \\
SC12 & 998  & 157.4/180 & 0.543 & 0.462 &  102.85 & $-0.09$ &  1398.7 &   26.6 &  0.33 & 0.0011 \\
SC15 & 1631 & 179.2/192 & 0.197 & 1.012 &  151.46 & $ 0.14$ &   578.9 &    7.8 &  0.59 & 0.0058 \\
SC16 & 1048 & 217.5/183 & 0.507 & 1.641 &  194.24 & $-0.21$ &   565.6 &   27.7 &  0.36 & 0.0023 \\
SC18 & 4924 & 156.2/194 & 0.292 & 1.084 &   18.94 & $-0.48$ &  1217.9 &   89.1 &  0.44 & 0.0007 \\
SC19 & 667  & 206.5/175 & 0.216 & 2.483 &   73.13 & $ 1.65$ &  1346.6 &   58.8 &  0.94 & 0.0027 \\
SC20 & 1793 & 133.9/225 & 0.017 & 2.917 &  144.03 & $ 1.25$ &   632.6 &    6.7 &  0.96 & 0.1003 \\
     &      & 134.2/225 & 0.501 & 0.355 &  319.24 & $-1.29$ &   615.9 &    6.3 &  0.95 & 0.1045 \\
SC20 & 3525 & 137.5/196 & 0.009 & 0.684 &   82.03 & $-0.02$ &  1336.6 &   77.8 &  0.14 & 0.0004 \\
     &      & 137.8/196 & 0.010 & 1.413 &   82.57 & $-0.03$ &  1331.7 &   67.6 &  0.16 & 0.0005 \\
SC20 & 4695 & 144.7/222 & 0.495 & 0.582 &  246.83 & $-0.08$ &   586.8 &  108.2 &  0.08 & 0.0002 \\
     &      & 145.0/222 & 0.822 & 0.861 &  205.26 & $ 0.27$ &   577.9 &   63.6 &  0.28 & 0.0006 \\
SC21 & 6195 & 206.1/223 & 0.543 & 1.135 &  104.33 & $-0.35$ &   601.9 &   56.3 &  0.46 & 0.0019 \\
SC30 & 4491 & 144.6/197 & 0.150 & 0.638 &  219.77 & $-0.11$ &  1448.5 &   74.0 &  0.23 & 0.0006 \\
SC31 & 1795 & 172.9/226 & 0.193 & 1.303 &   83.01 & $ 0.51$ &  1310.3 &   18.9 &  0.75 & 0.0046 \\
SC31 & 3205 & 228.7/225 & 0.106 & 0.750 &  347.39 & $-0.06$ &  1271.9 &   78.3 &  0.17 & 0.0003 \\
SC32 & 4683 & 150.8/198 & 0.176 & 1.274 &   54.72 & $-0.19$ &  1146.4 &  243.6 &  0.04 & 0.0001 \\
SC35 & 2526 & 140.0/169 & 0.785 & 0.442 &  304.50 & $-0.07$ &   682.5 &  105.9 &  0.35 & 0.0004 \\
     &      & 140.1/169 & 1.000 & 2.818 &   29.70 & $ 0.71$ &   821.0 &  119.9 &  0.54 & 0.0005 \\
SC36 & 4030 & 182.3/180 & 0.292 & 1.462 &   63.14 & $ 0.46$ &  1300.4 &   94.7 &  0.45 & 0.0007 \\ 
\noalign{\vskip3pt}
\hline}
 
The source radius $r_s$ in Einstein units, which we give in the last column is 
not a fitted parameter. It is only estimated and used for the self consistency 
check. 

We concentrate mostly on the physical parameters of the binary ($q$ and $d$). 
Examining the plots one can see that not all events can be well constrained 
and for some the $1\sigma$ confidence regions are rather large and irregular. 

\subsection{Distribution of Mass Ratios and Separations}
There are 12 caustic crossing binary lens events among the total of 18. The 
conditional probability of a binary lens to cause a caustic crossing event is 
well defined (Mao and Paczy\'nski 1991). It is proportional to the angle 
averaged width of caustic measured in the direction perpendicular to the 
source path and can be calculated for any binary lens as $w(\lg q,\lg d)$, 
where we use the dependence on logarithms as more convenient. The most 
efficient lenses have ${q\approx1}$ and ${d\approx1}$. 

The rate of observed caustic crossing events caused by lenses with given 
parameters $P_{\rm BCC}(\lg q,\lg d)$ also depends on the distribution of 
binary stars parameters $P_{\rm bin}(\lg q,\lg d)$. Following Mao and 
Paczy\'nski (1991) we assume, that the uniform distribution of binary systems in 
the logarithm of orbital period, which holds over six decades (Abt 1983) 
translates into uniform distribution in the logarithm of semimajor axis, and 
consequently in $\lg d$. Under this assumption the distribution of parameters 
among observed events is given by: 
$$P_{\rm BCC}(\lg q,\lg d)\sim P_{\rm bin}(\lg q) w(\lg q,\lg d)\eqno(4)$$
and, in principle, may be used to constrain $P_{\rm bin}(\lg q)$. The binary 
systems database has been investigated by Trimble (1990), who shows that the 
mass ratio distribution is uniform in $\lg q$. This dependence is not valid 
for small $q$  where the binaries are less abundant. We consider distributions 
of the form: 
$$P_{\rm bin}(\lg q)\sim 
\begin{cases}
q/q_0   &     q \le q_0\\
  1     &     q \ge q_0
\end{cases}
\eqno(5)$$
using ${0.03\le q_0\le 0.3}$ which defines the location of the power law 
break. Using the above formulae one can find the integral probability 
distribution for the mass ratio and separation among the binaries causing BCC 
events. It is not obvious how to model the probability distribution for the 
cases without caustic crossings. We assume that our calculation, which 
includes only the caustic crossing events, may be applied to a sample 
including both caustic crossing and cusp approach events. The comparison of 
the theoretical predictions with the distribution of the binary lens sample, 
excluding the event 1998-BUL-28, which shows no sharp feature in the light 
curve. is shown in Fig.~1. 
\begin{figure}[htb]
\includegraphics[width=6.0cm]{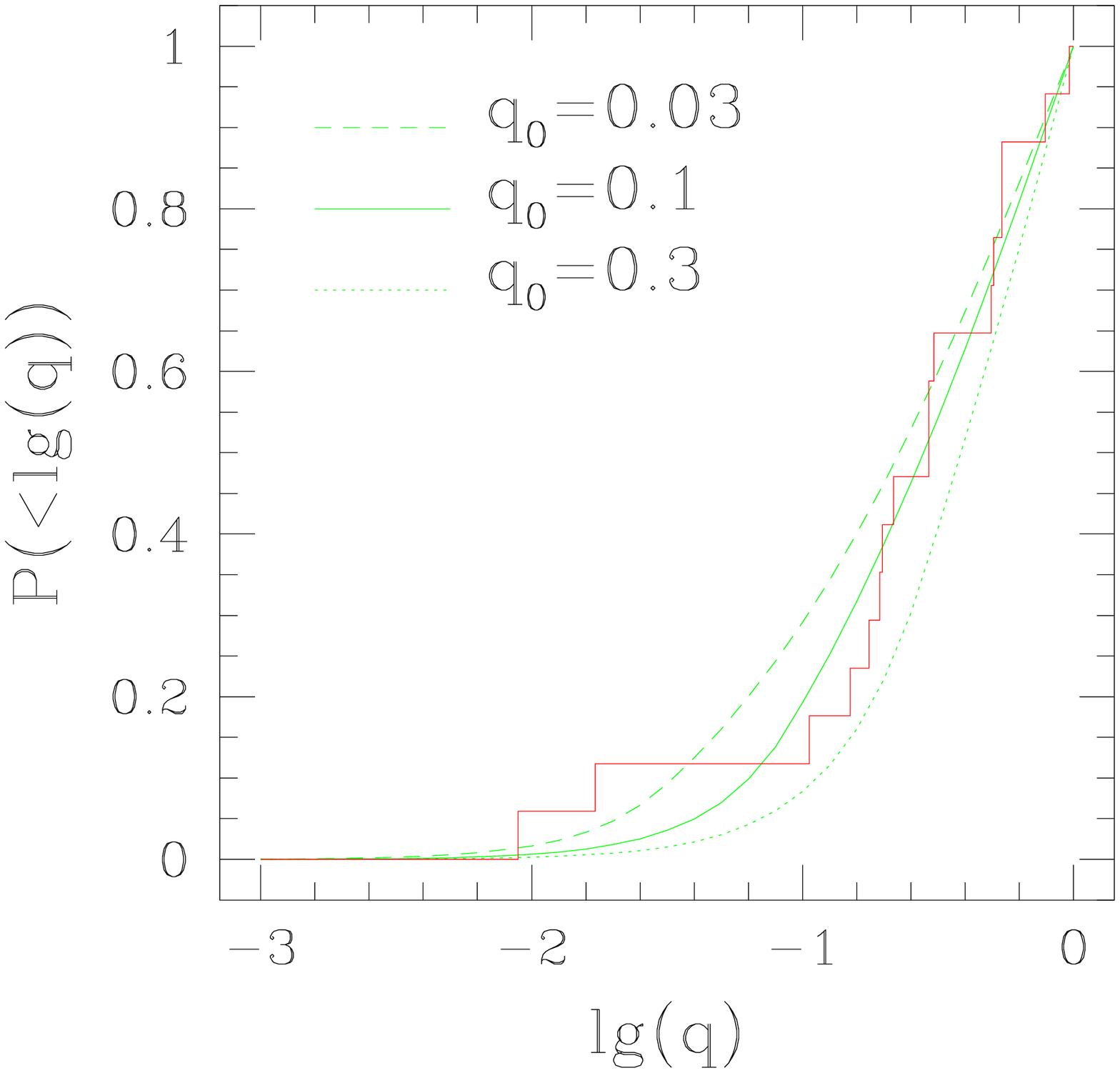}\hfill
\includegraphics[width=6.0cm]{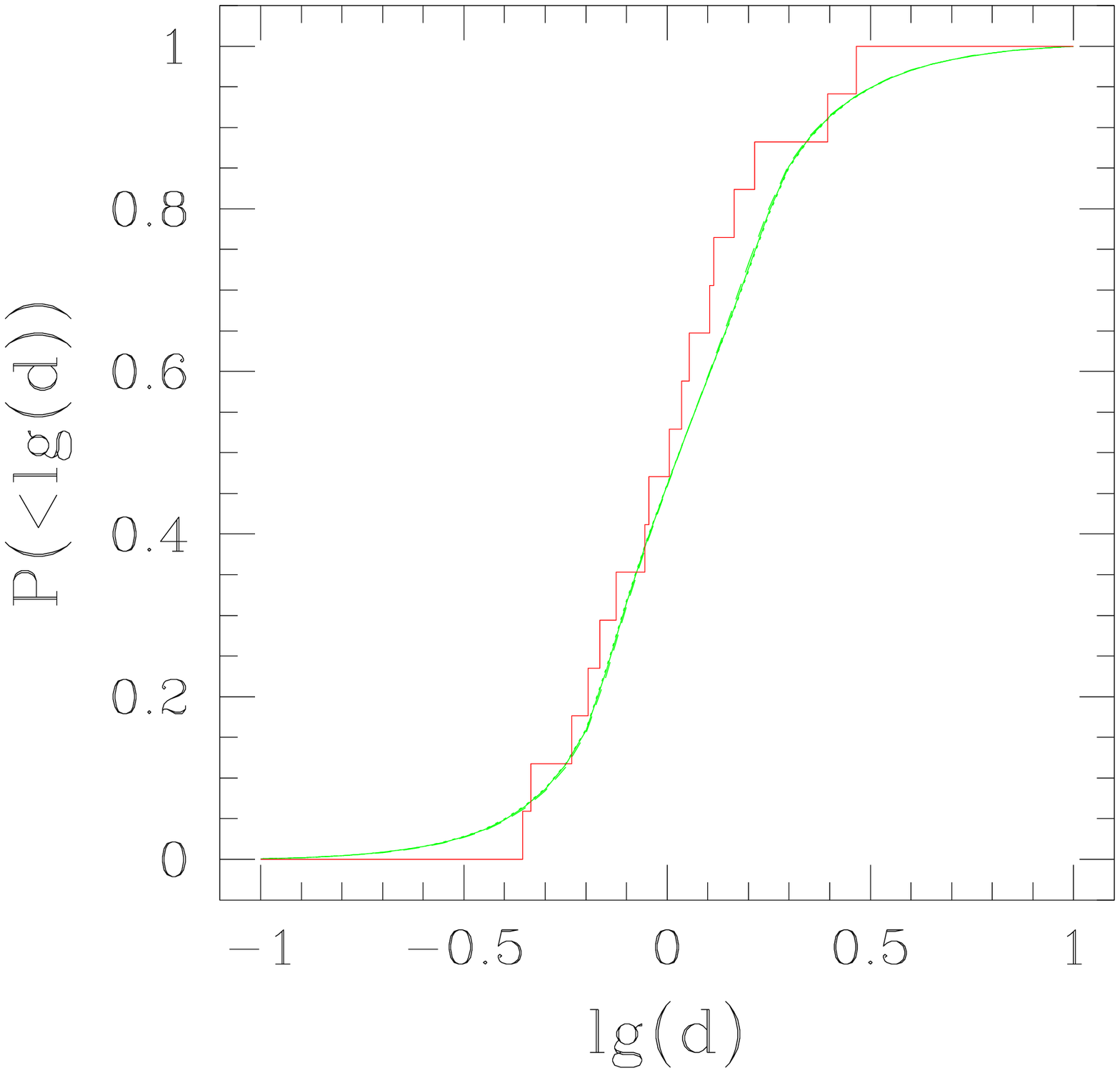}
\FigCap{Integral probability distributions for the mass ratio (left) and the 
binary separation (right) as compared with the binary lens sample. Curves 
corresponding to the same limiting mass ratio values (${q_0=0.03{-}0.3}$) are 
shown on both plots, but the curves on the right panel are indistinguishable. 
(See the text for details on binary star population modeling.)} 
\end{figure}

Our binary sample consists of the best models for each event. There are two 
cases with extreme mass ratios: SC20\_1793 and SC20\_3525. Both are cusp 
approach events. The first of the cases has also many other fits within the 
$1\sigma$ confidence limits with ${q>0.1}$. The other case is a more serious 
candidate for an extreme mass ratio binary lens, but models with ${q>0.1}$ are 
present within $2\sigma$ confidence limits. 

The mass ratio and separation distributions of our sample (excluding possible 
extreme cases) is in rough agreement with theoretical expectations. The number 
of cases is low, so it is too early to make definite conclusions regarding the 
distribution of parameters among the binary stars. 

\subsection{Distribution of Einstein Radius Crossing Time}
The Einstein radius crossing time is a physical parameter depending on the 
distances, lens mass and relative velocity, which can be found both for single 
and binary lenses. We show the histogram of this quantity for our sample of 
binary lens candidates in Fig.~2. 
\begin{figure}[p]
\centerline{\includegraphics[width=8.5cm]{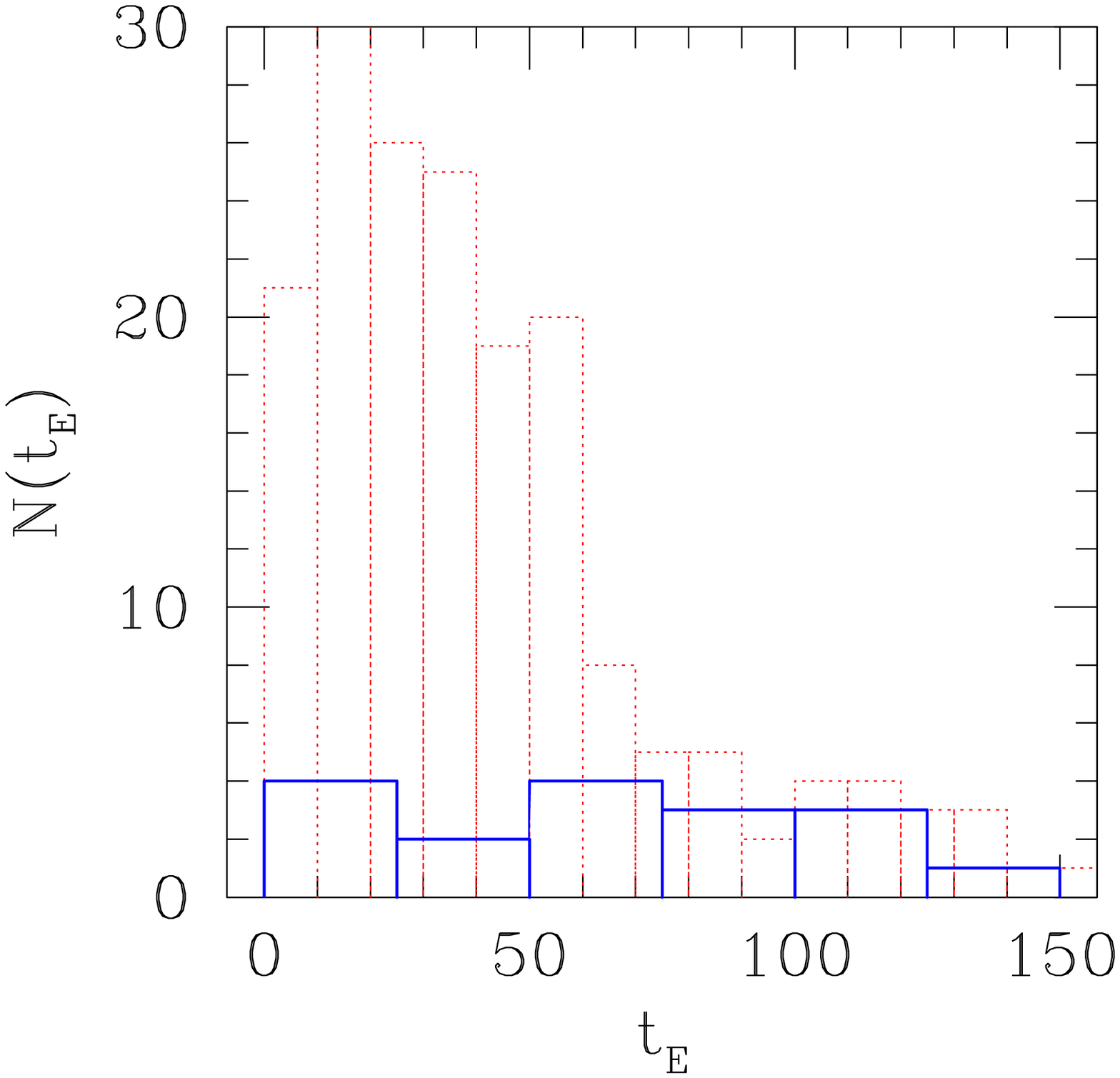}}
\vskip-5mm
\FigCap{Distribution of the Einstein radius crossing time $t_{\rm E}$ for the 
sample of binary lenses (solid line) as compared with the sample of single 
lenses (dotted) selected from the transient database using the same 
criteria.} 
\vskip10mm
\centerline{\includegraphics[width=9cm]{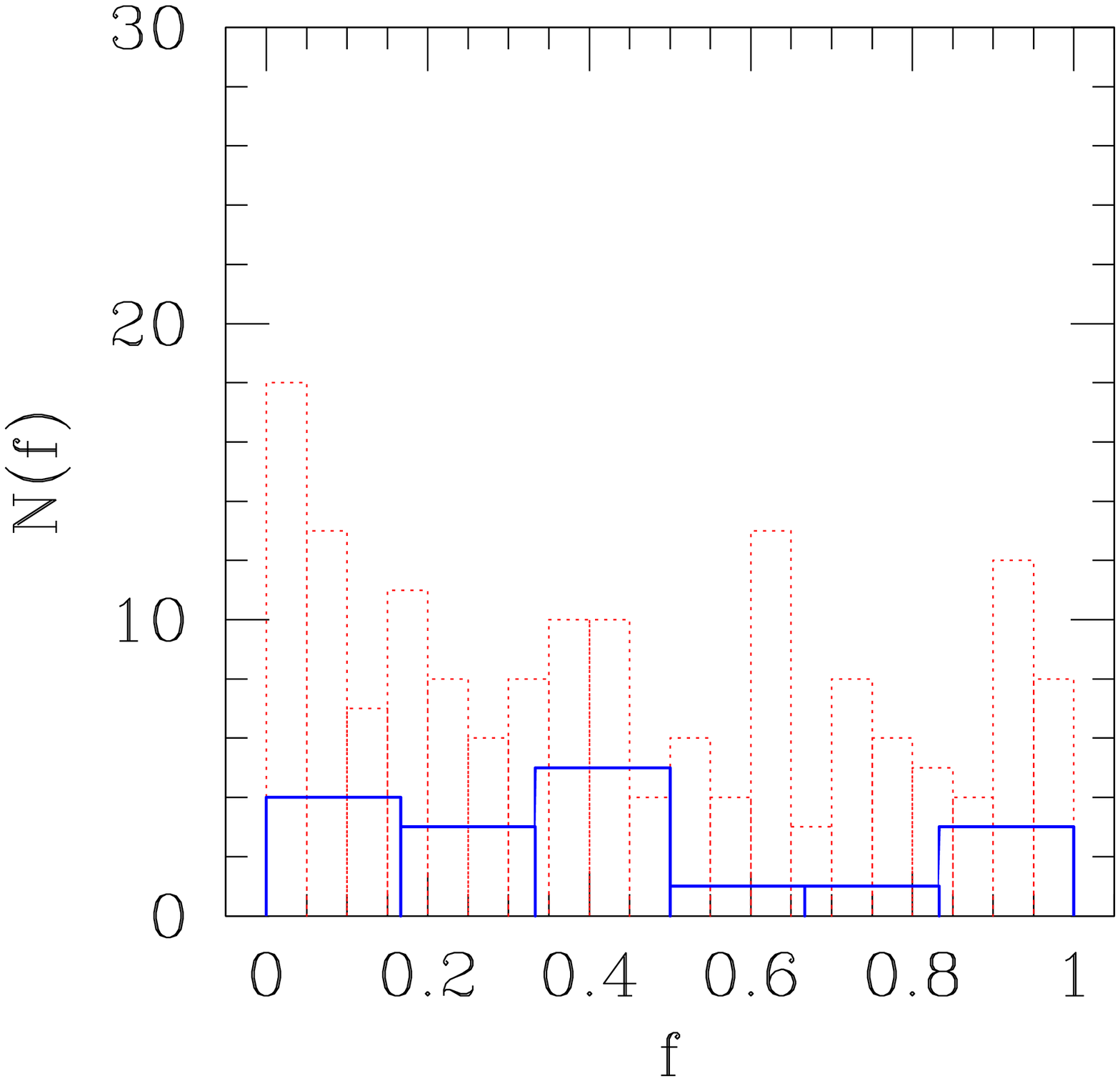}}
\vskip-7mm
\FigCap{Distribution of the parameter $f$ defining which part of the base flux 
belongs to the source. (Part proportional to ${1-f}$ belongs to stars close to 
the source and to the lens \ie the blends.) The histogram for binary lenses 
(solid) is compared with the histogram for our single lens sample (dotted).} 
\end{figure}

The statistics is poor, but the over-representation of the long lasting events, 
as compared with our single lens sample  is apparent. The binary lenses should 
be twice as massive on average, which translates to the $\sqrt2$ factor in 
Einstein time. This is, however, not sufficient to explain the difference 
between the distributions. 

\subsection{The Blending Parameter $f$}
This parameter, showing the fraction of the base flux belonging to the lensed 
star (while the rest is blended) can be fitted with greater confidence than 
for single lenses. We force this parameter to belong to the region ${0.01\le f 
\le0.99}$. Only one of our models lies on its upper boundary. (Sometimes one 
can obtain unphysical solutions with ${f>1}$ and low $\chi^2$ value which 
represent blends of negative fluxes. Other rather unlikely models can have ${f 
\rightarrow0}$ and very high amplification -- usually related to crossing of a 
small caustic. None of our models falls into the second category.) We show a 
histogram of $f$ values in Fig.~3. 

For better comparison of single and binary lens models we show their positions 
on $\lg(t_{\rm E})$--$f$ diagram. We use the convention employed also by 
Wo\'zniak \etal (2001), which preserves single lens fits with unphysical values 
of blending parameter (${f>1}$) implying the negative flux of light from the 
blend. (If forced to ${f\le1}$, these models cluster on the boundary of 
permitted region.) Another unphysical region on the diagram consists of points 
with ${f\rightarrow0}$ and high $t_{\rm E}$. The presence of single lens 
models in the unphysical regions of the diagram is explained by Wo\'zniak and 
Paczy\'nski (1997) and is related to the degeneracy among fitted  parameters in 
certain situations. 
\begin{figure}[htb]
\centerline{\includegraphics[width=8.5cm]{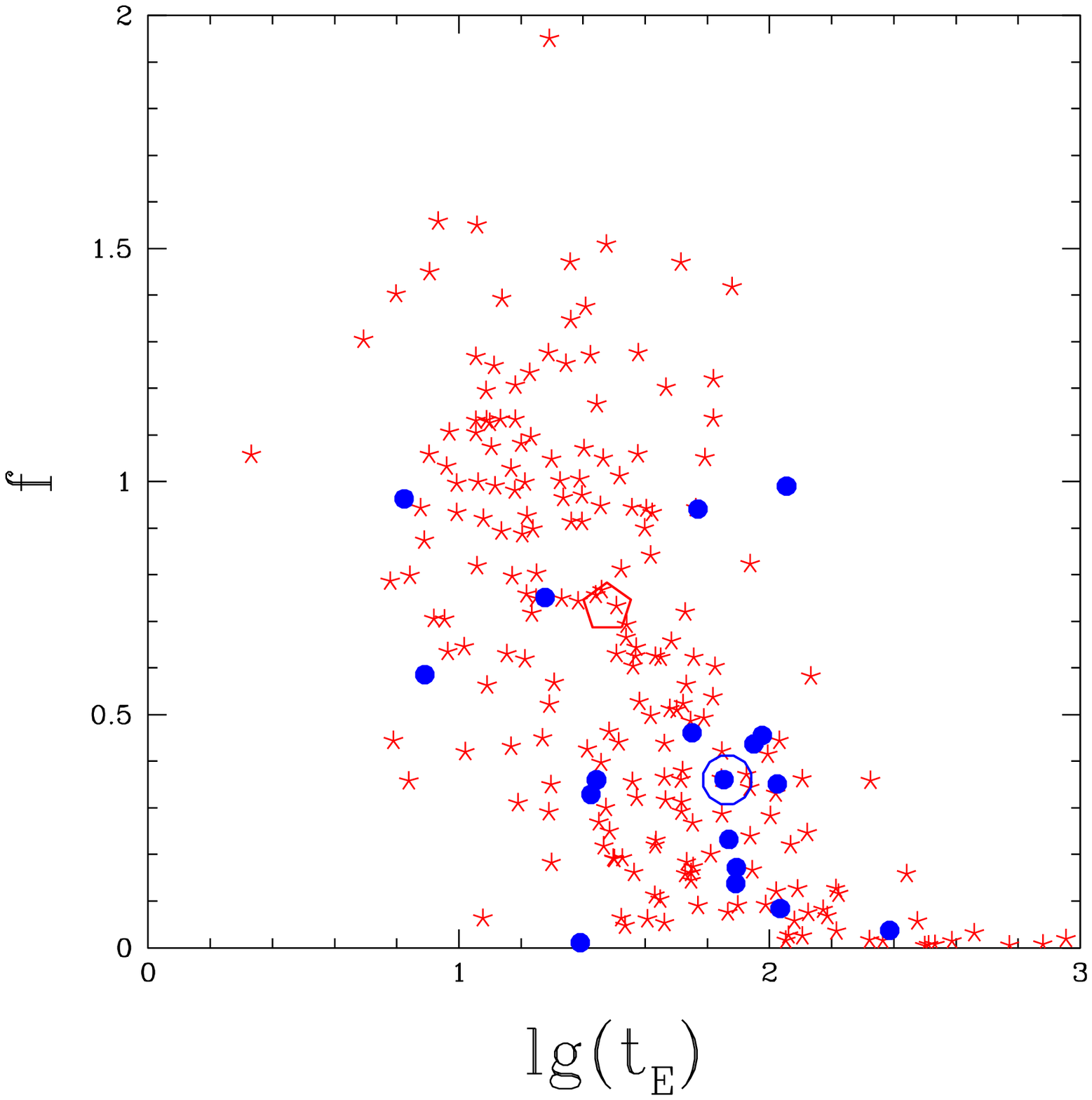}}
\vskip-7mm
\FigCap{Positions of binary(dots) and single (stars) lens models on the 
$\lg(t_{\rm E})$--$f$ diagram. The median points for binaries (open circle) 
and singles (open pentagon) are also shown.} 
\end{figure}

In Fig.~4 we show the single and binary lens models belonging to our sample. 
The median points for the two kinds of lenses are different: the binaries have 
systematically longer duration (73\upd versus 30\upd for singles) and are more 
blended (${f_{\rm med}=0.36}$ versus 0.62 for singles). It is interesting to 
point out that for all of the 520 single lenses of Wo\'zniak \etal (2001) the 
median duration is even shorter (23\upd) and the blending less important 
(${f_{\rm med}=0.88}$). The simple rejection of all models with ${f>1}$ gives 
almost equal values of median blending parameters (${f_{\rm med}\approx0.37}$) 
for binaries, singles from our sample and singles from Wo\'zniak \etal (2001). 
The median Einstein times become longer for ${f<1}$ subsamples of single lens 
models, but still shorter than for binaries (42\upd/33\upd versus 73\upd). 
There is no simple criterion for removing models with ${f\rightarrow0}$ and we 
do not pursue this issue. 

\subsection{Contribution of Binary Lenses to Microlensing Optical Depth}
Gravitational microlensing is detected because of the temporary presence of an 
extra flux of light caused by source amplification. The detailed modeling of 
the detection process is a separate issue, but it is clear that both the 
duration of the event and the strength of the lens play here important roles. 
The customary definition of lensing event (Paczy\'nski 1996) requires the object 
to be amplified at least 1.34 times, which corresponds to the source position 
inside the Einstein ring. Thus the contribution of a single event to the 
optical depth is proportional to the time it spends above the threshold 
amplification. For a blended source the observed amplification is not so 
simply related to the lens position. In DIA photometry it is rather the flux 
difference, which is important in discovering lensing events. We estimate the 
relative contribution of binary and single events to the microlensing optical 
depth comparing the extra light related to the two kinds of phenomena. The 
extra light corresponding to a particular event is given as: 
$$a_i=F_{\rm s}\int(A_i(t)-1)\,{\rm d}t\eqno(6)$$
where index $i$ enumerates the events. Comparing the whole extra light of our 
binary lenses with the extra light of single lenses we obtain: 
$${\frac{\sum_{\rm bin}a_i}{\sum_{\rm sin}a_i}}=0.03\eqno(7)$$
which suggests a 3\% contribution of binary lenses to optical depth. The base 
fluxes of objects belonging to our sample which undergo single lensing are 
systematically higher than for objects undergoing binary lensing. This may be 
the result of our selection criteria. The effect diminishes the importance of 
binary lenses. It is possible that sources amplified by binary lenses are on 
average fainter as compared to sources undergoing lensing by single masses for 
a wide class of selection criteria. We check also other possibilities using 
our sample of single and binary lenses but changing the population of sources. 
This should be treated as a thought experiment which serves only to estimate 
the likely systematic errors of the above result -- it is not guaranteed that 
for a different population of sources the same population of lenses would be 
discovered. 

In the first experiment we assume that the base flux ($F_{\rm base}$) of all 
objects is the same. We keep unchanged the other parameters of the lens models 
(including $f$). Calculating the ratio of extra light contributed by binary 
and single lenses we obtain the value 0.11. In the second experiment we assume 
that the source fluxes ($F_{\rm s}$) are the same for all objects. This gives 
0.05 for the calculated ratio. 

The preliminary analysis of our sample shows that the recognizable binary 
lenses make about 8\% of all lensing events in excellent agreement with Mao 
and Paczy\'nski (1991). Estimating the contribution to optical depth by binaries 
is a more difficult issue. Using three different approaches we obtain the 
values between 3\% and 11\%. 

\section{Discussion}
Our fitting procedure is based on two important simplifications. First we use 
mostly a point source model while calculating the source magnification. We 
make some self consistency checks using also extended source models, but we 
are not able to fit its size. The effects of source size are important during 
caustic crossing and/or cusp approaches. Some of the models would have 
probably been changed if the data were better sampled. Only the event 
SC36\_4030 (1999-BUL-11), for which there are many observations  of the second 
caustic crossing might be used to constrain the source size. The observations 
near the caustic are, however, marked as not to be trusted by DIA photometry 
pipeline and we do not include them. The second simplification is the fact 
that we neglect all effects of the Earth motion, rotation of the binary, and 
possible acceleration of the source. In fact we have attempted fits including 
the effects of parallax/acceleration, but finally we have decided not to use 
them. Formally for long lasting events the Earth motion should be included. In 
the case of single lens events, the Einstein time is simply related to the 
duration of strong amplification. In the case of a binary this is not 
necessarily so: the passage through or near a small caustic, which implies 
strong lensing, may last much shorter than the Einstein time. The observations 
of only weekly magnified source are not discriminating between the models. 
Thus, even for formally ``long'' events it is not easy to measure the effects 
of changes in the relative velocity of the source. Introducing extra 
parameters describing the relative acceleration of the source may serve as to 
lower the $\chi^2$ value appreciably (at least for some events), but the extra 
parameters cannot be well constrained. 

The estimates of the caustic size which we use for self consistency checks may 
also be used to place limits on possible lens mass and luminosity. In some 
cases the resulting lens brightness is too high. We do not consider this a 
serious problem, since the relative velocity of the lens may be substantially 
different from the ``typical'' value used in the estimates. 

For two events (SC20\_1793 and SC20\_3525) the formally best binary lens 
models have extreme mass ratios (${q<0.02}$), which may suggest they contain 
brown dwarfs or Jupiter-like objects. The same events can be, however, modeled 
using binary lens models with much higher mass ratio, which cannot be 
rejected with high confidence. Some parts of the parameter confidence region 
also reach low mass ratio values (${q\approx10^{-2}}$) for SC35\_2526, but 
this is a case of very weakly constrained solutions. Statistically acceptable 
models may belong to a wide region of mass ratios. Thus we cannot conclude a 
presence of a planetary system among binary lens candidates. 

The distribution of Einstein radius crossing times among the models requires 
further investigation. The occurrence of long lasting events is certainly more 
likely in our sample as compared to single lens sample. The same effect but 
to lesser extent is also present in Alcock \etal (2000) sample of binary lens 
candidates. Since the recognition of binary events requires longer baselines, 
there is certainly some selection effect involved. The essential part of the 
light curves -- caustic crossings or cusp approach (approaches) may last only 
a small fraction of the Einstein time. That puts some lower limit on the 
inter-caustic time, depending on the frequency of observations. With 
observations made once a day we are biased toward long lasting events. 

\subsection{Comparison with Previous Study}
We are using DIA processed observations of OGLE-II in the direction of 
Galactic bulge made in seasons 1997--1999 (Wo\'zniak \etal 2001). The same 
observational material, but processed differently, has been analyzed by 
Udalski \etal (2000), who explicitly mark the binary lens candidates in their 
Table~2. We list all their candidates in our Table~3, and show their status 
according to the present study. 
\MakeTable{lccccl}{12.5cm}{Comparison with previous study}
{\hline
\noalign{\vskip3pt}
\multicolumn{4}{c}{Identification}& \multicolumn{2}{c}{Remarks} \\
Field & OGLE & EWS & DIA & Udalski & This study \\ 
\noalign{\vskip3pt}
\hline
\noalign{\vskip3pt}
BUL\_SC3  & 541151 &       --     & 5453 & BIN? & SIN \\
BUL\_SC3  & 590098 &       --     & 7783 & BCC  & not a lens \\
BUL\_SC3  & 792295 &  1998-BUL-28 & 7390 & BCC  & nCC \\
BUL\_SC15 & 373196 &       --     & 1631 & BCC  & BCC \\
BUL\_SC16 & 32304  &       --     & 1048 & BCC  & BCC \\
BUL\_SC19 & 26606  &  1999-BUL-23 &  669 & BCC  & BCC \\
BUL\_SC19 & 64524  &       --     & ---- & BCC  & not a transient \\
BUL\_SC20 & 391296 &       --     & 5451 & BIN? & SIN \\
BUL\_SC20 & 395103 &       --     & 5875 & BIN  & not selected  \\
BUL\_SC20 & 708586 &  1999-BUL-25 & 3525 & BCC  & nCC \\
BUL\_SC21 & 833776 &       --     & 6195 & BCC  & BCC \\
BUL\_SC23 & 282632 &       --     & 2041 & BIN? & not a lens \\
BUL\_SC30 & 352272 &       --     & ---- & BCC  & not a transient \\
BUL\_SC31 & 293442 &  1999-BUL-17 & 1795 & BCC  & BCC \\
BUL\_SC33 & 476067 &  1999-BUL-42 & ---- & BCC  & not a transient \\
BUL\_SC34 & 87132  &       --     & 2535 & BCC? & SIN \\
BUL\_SC35 & 305604 &       --     & 2526 & BCC  & BCC  \\
BUL\_SC36 & 336761 &  1999-BUL-11 & 4030 & BCC  & BCC \\
BUL\_SC39 & 259656 &  1999-BUL-45 & 1405 & DLS  & SIN? \\
BUL\_SC39 & 378193 &       --     & 5900 & BIN  & SIN? \\
BUL\_SC40 & 434222 &  1999-BUL-19 & ---- & DLS  & not a transient \\ 
\noalign{\vskip3pt}
\hline
\noalign{\vskip3pt}
\multicolumn{6}{p{9.5cm}}{Note: The event identification is given in columns 
1--4. We repeat the remarks of Udalski \etal (2000) in column~5, where the 
acronyms have the following meaning: BCC -- binary lens with caustic crossing, 
BIN -- possible binary lens, and DLS -- binary lens or double source star. In 
column 6 we give our own remarks using also: SIN -- single lens, and nCC -- 
binary lens with no caustic crossing. The events not in the transient sample 
are marked as ``not a transient'', and events rejected after modeling as binary 
lenses -- ``not a lens''.}} 

Four of the Udalski \etal (2000) binary lens candidates (see Table~3) are not 
in the transient database of Wo\'zniak \etal (2001), and another one has not 
passed our selection criteria. Two of the events were also on the list of our 
27 candidates, but have been rejected after the attempt to fit them with 
binary lens model. In five cases we believe the events are due to single lens 
rather than binary. (Three cases are strong: the DIA photometry does not show 
the irregularities of the light curves near maximum, which are present on the 
plots in the Udalski \etal (2000) catalog. Two other cases are examples of 
noisy light curves and we treat them as poor quality single lens cases.) In 
nine cases we agree that the events are due to binary lenses, but in two of 
them we think there is no caustic crossing. 

The comparison of our sample with the list of 520 microlensing events of 
Wo\'zniak \etal (2001) is also due, because in both cases the same database has 
been used. Our approach differs since we apply stronger selection criteria to 
have a better chance of distinguish the binary lens candidates. The possible 
binary lens events are not marked in the other study, so a direct comparison 
is not possible. Of our original 27 binary lens candidates only 3 are missing 
on the Wo\'zniak \etal (2001) list. After attempting to model the 27 candidate 
events as binary lenses, we reject 9 of them, including the three cases 
mentioned above. According to our study, the rejected events are not due to 
microlensing, and this includes 6 cases from the Wo\'zniak \etal (2001) list; at 
the same time the list includes all 18 of our final binary lens candidates. 
The single lenses have not been studied extensively. Two events, which we 
classify as single microlensing are missing in the Wo\'zniak \etal (2001) sample 
-- these are rather poor cases of microlensed light curves. Two events, which 
we mark as ``other objects'' belong to the list of Wo\'zniak \etal (2001). 

The event SC\_21 6195 has been considered also by Alcock \etal (2000) (their 
symbol: 97-BLG-d2.) Both models are qualitatively similar, representing 
intermediate separation binaries and source trajectory approaching one cusp 
and then crossing the caustic twice near another cusp. Model parameters differ 
from 7\% (impact parameter) to 40\% (mass ratio). The parameter defining 
blending can not be compared directly because of different filters used by 
OGLE and MACHO teams. 

\Acknow{We thank Bohdan Paczy\'nski for many helpful discussions. Special thanks 
are due to Shude Mao for the permission of using his binary lens modeling 
software, to Przemek Wo\'zniak for the permission to use DIA photometry 
transient database before its publication, and to OGLE team for their generous 
early release of data. This work was supported in part by the Polish KBN grant 
2-P03D-013-16 and the NSF grant AST 98-20314.}

\newpage
\begin{figure}[p]
\vspace{25.40cm}
\includegraphics{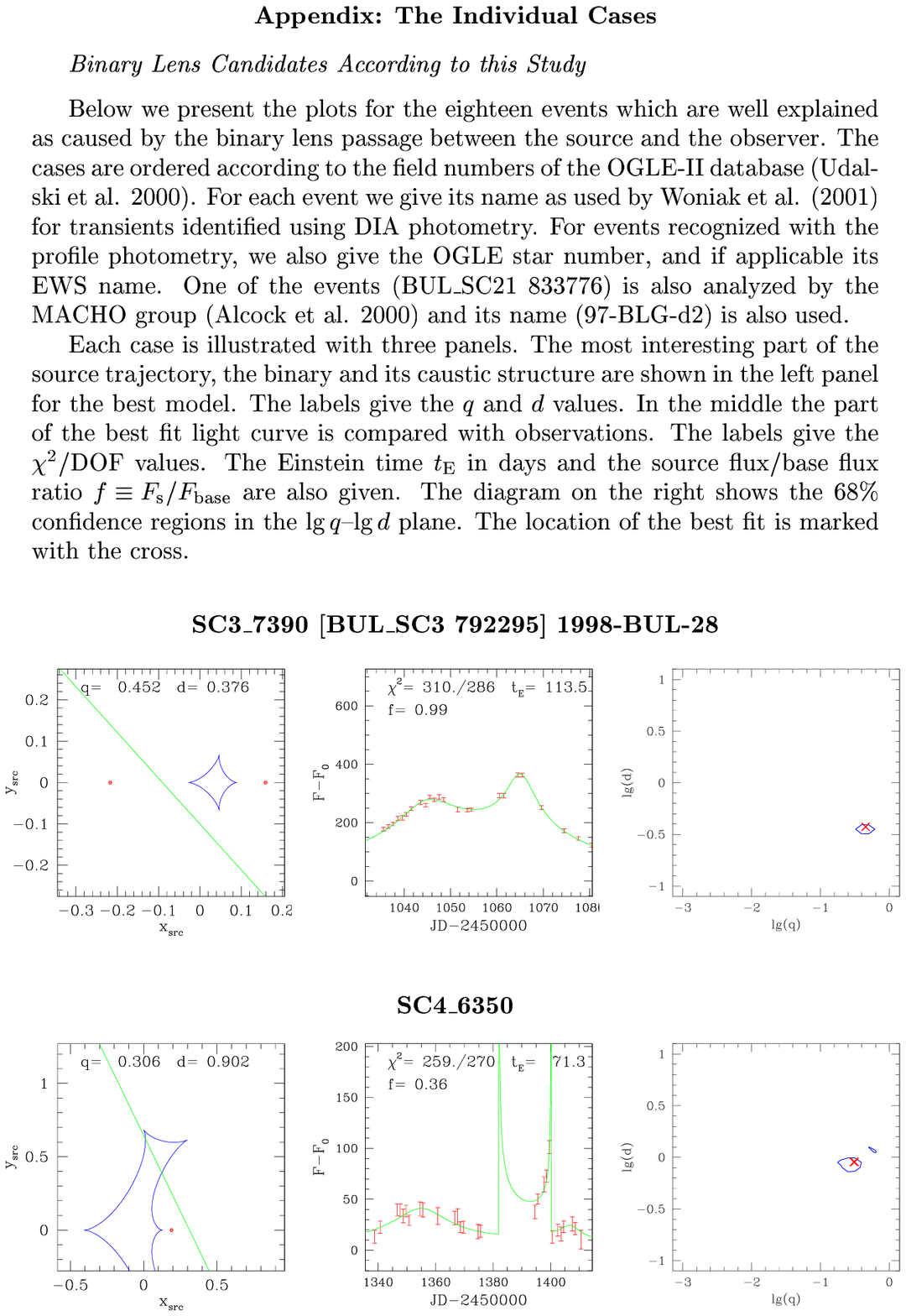}
\end{figure}

\begin{figure}[p]
\vspace{25.40cm}
\includegraphics{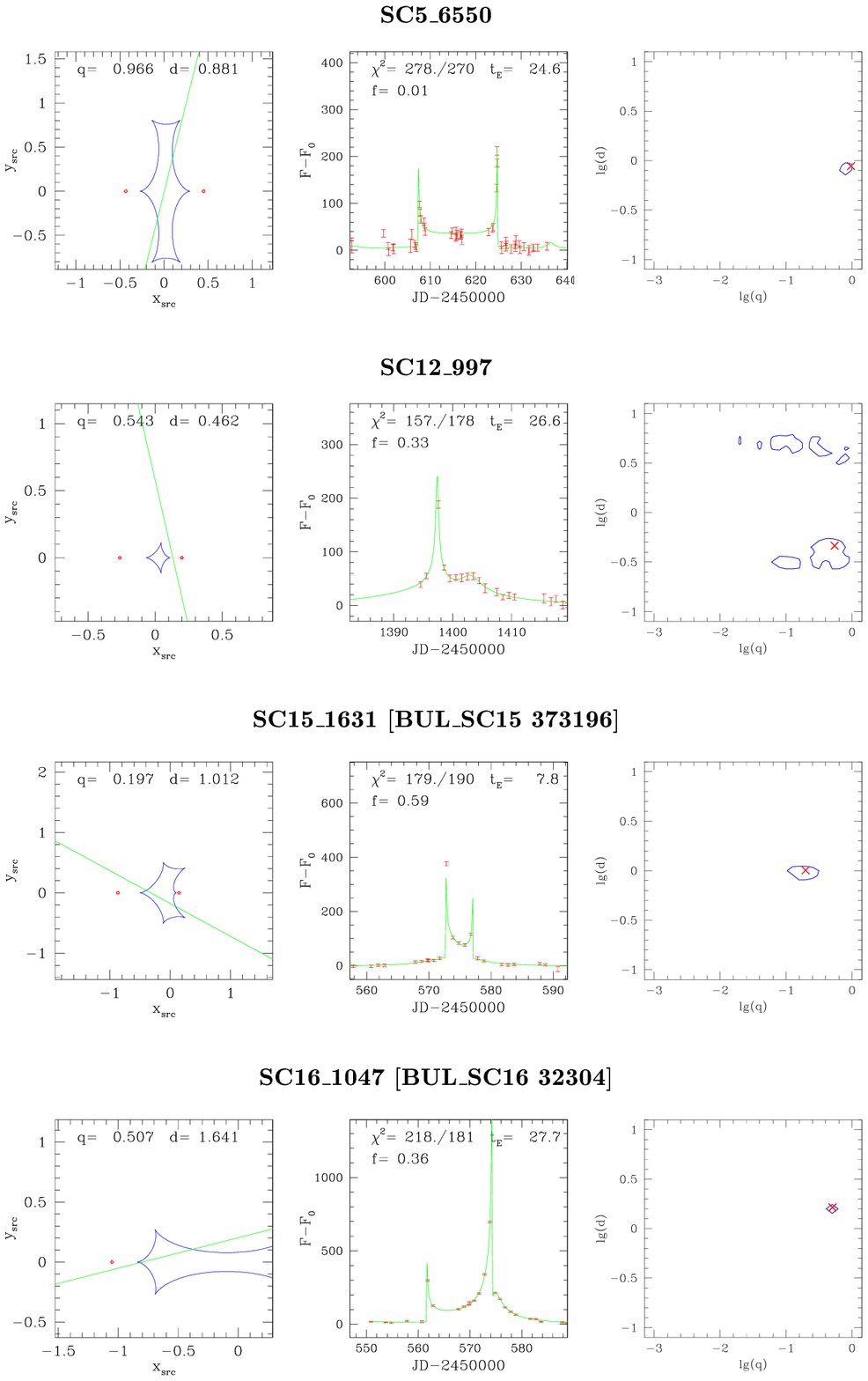}
\end{figure}

\begin{figure}[p]
\vspace{25.40cm}
\includegraphics{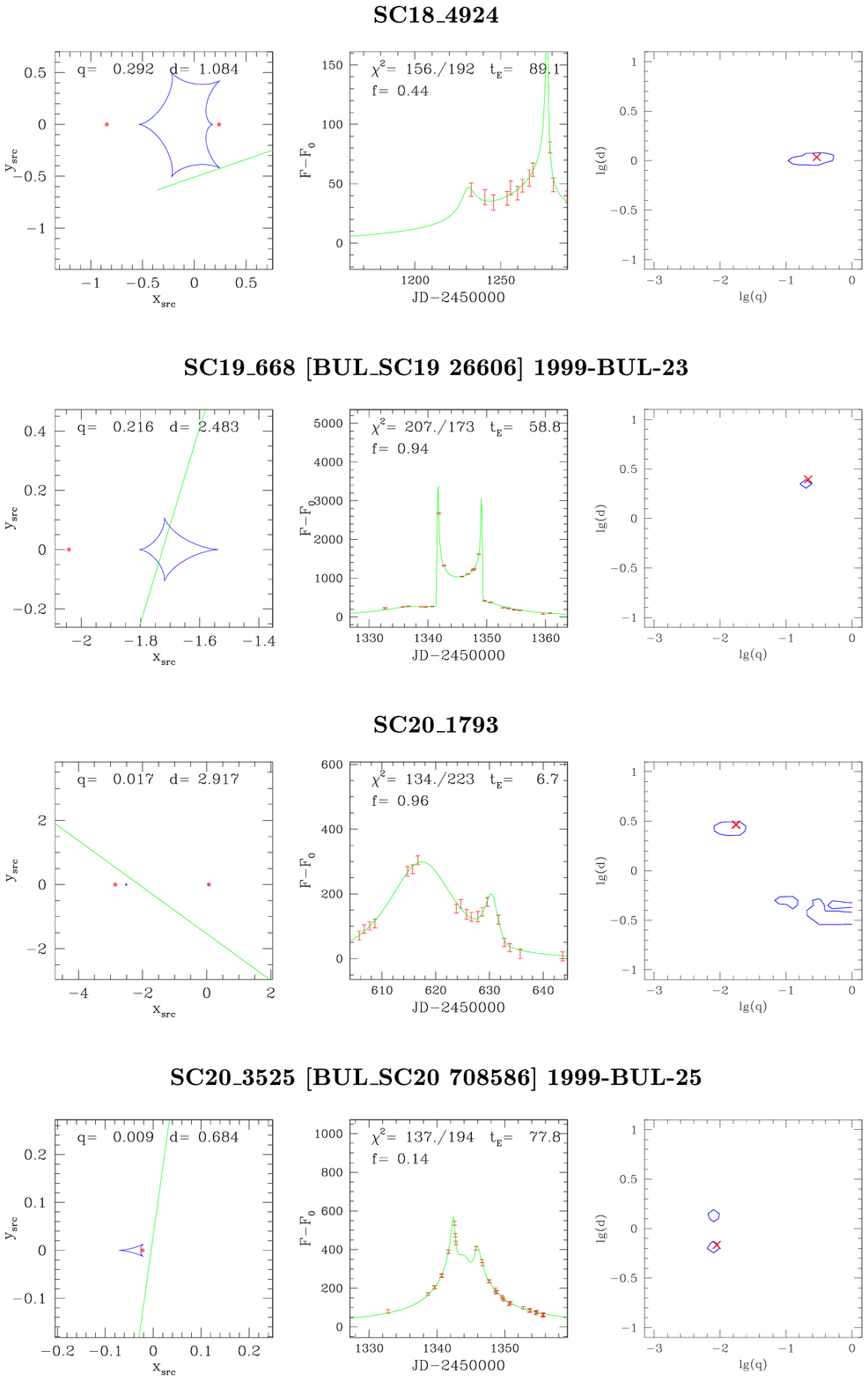}
\end{figure}

\begin{figure}[p]
\vspace{25.40cm}
\includegraphics{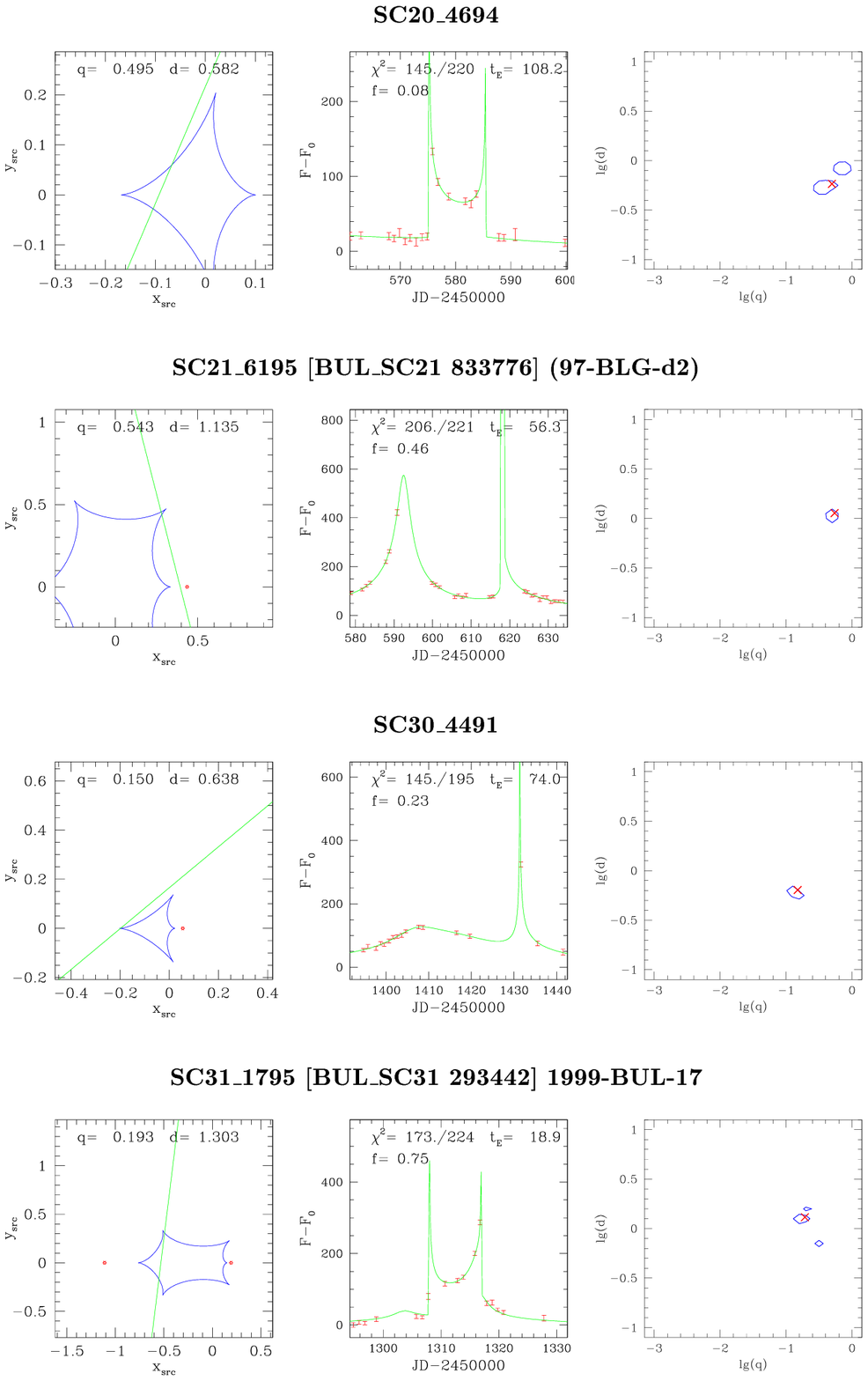}
\end{figure}

\begin{figure}[p]
\vspace{25.40cm}
\includegraphics{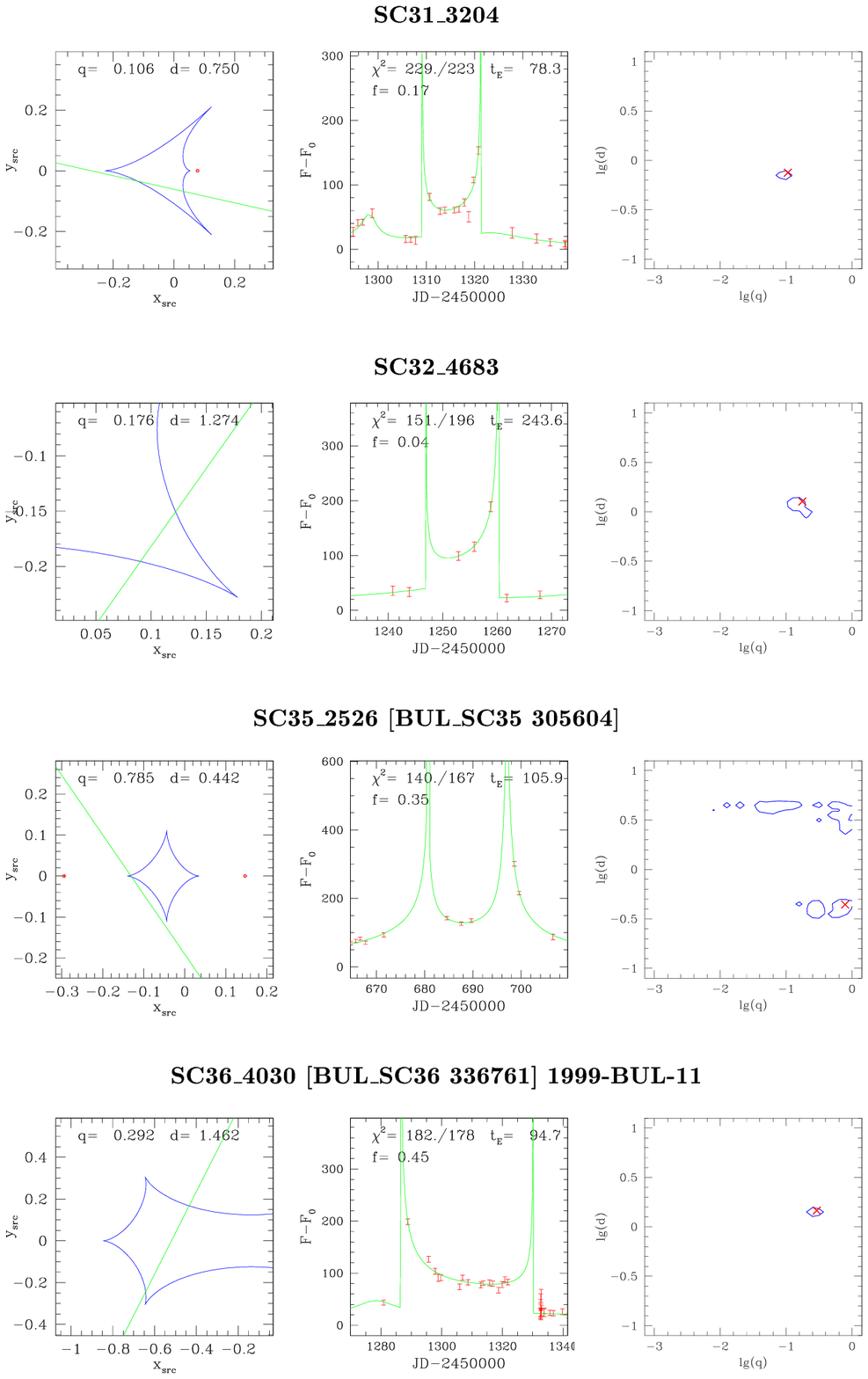}
\end{figure}

\begin{figure}[p]
\vspace{25.40cm}
\includegraphics{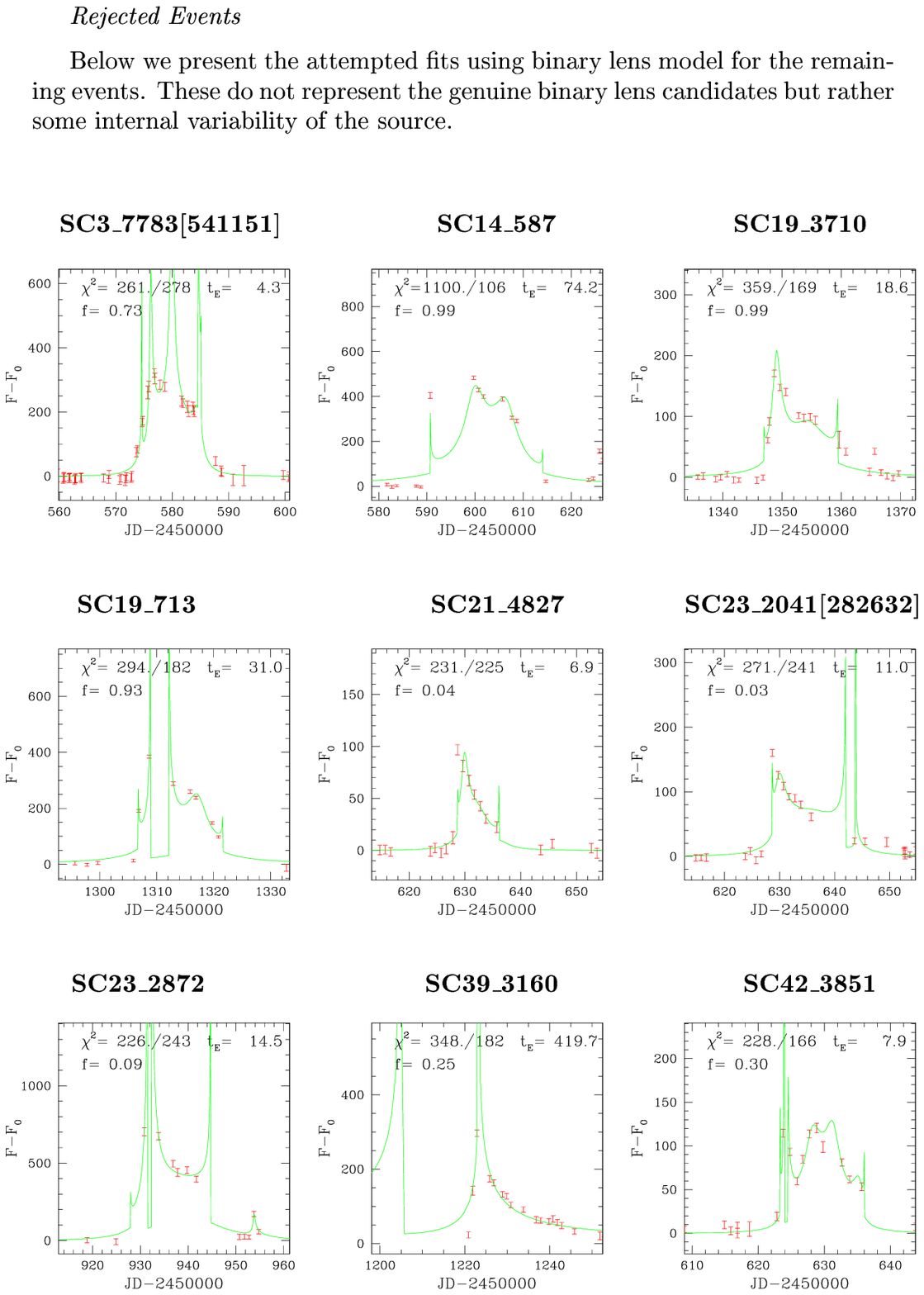}
\end{figure}

\end{document}